\documentclass[aps]{revtex4}

\begin{document}
\title{Colour-$SU(3)$-Ginzburg-Landau Effective Potential 
for Order Parameter with $3 \times 3$ Symmetry II : \\ 
 A Complete Classification of Phase Diagrams}
\author{E. Nakano, T. Suzuki and H. Yabu }
\affiliation{Department of Physics, Tokyo Metropolitan University, 
         1-1 Minami-Ohsawa, Hachioji, Tokyo 192-0397, Japan}
%
%
\begin{abstract}
Phase structure is studied for 
a system which has symmetry group $SU(3)$ and 
is described by $SU(3)$-${\bf 3 \times 3}$ order parameter. 
The study is rested on $SU(3)$-Ginzburg-Landau effective potential 
constructed as a preliminary. 
\end{abstract}
\maketitle
%
\section{Introduction}
%
In the previous paper \cite{NYS}, 
we constructed the Ginzburg-Landau effective potential 
for the system which has 
$SU(3)$ symmetry and is described by the order parameter 
that transform as the $(3, 3)$ representation under the $SU(3)$ group.
We also obtained the stability condition on the coefficients 
which are assigned to each $SU(3)$-invariant in the effective potential.

This work was motivated by the theoretical studies for 
the quark Cooper-pair condensation in a high-density quark matter, 
which is well characterised 
by the spontaneous breaking of colour $SU(3)$ symmetry \cite{CSC},  
and also by the academic interest as an extension of 
the models with the spontaneous symmetry breaking built-in, 
e.g. Ginzburg-Landau models of superfluid phase of ${}^3$He \cite{VOL} 
or Goldstone model as a field theoretical one \cite{GS}.

Corresponding to the irreducible representations of 
$(3, 3)$ symmetric order parameter, 
there are four kinds of the phases emerging in our model. 
To complete the analysis of the effective potential 
the phase diagram must be obtained for whole the coefficients region
on which the stability condition were already imposed \cite{NYS}.
For this purpose, 
we first analyse the possible region of each phase on the coefficients and then 
straightforwardly compare the magnitudes of these potentials 
for whole region of the coefficients.
In general, 
the magnitude of the effective potential depends not only 
on the scale of the order parameter but also on its angular parameters.

In the II section, 
we briefly review our model for the system mentioned above.
The III section describes a procedure to derive the phase diagram.
The IV section gives the results and conclusions.
%
\section{THE SYSTEM WITH $SU(3)$-${\bf 3 \times 3}$ ORDER PARAMETER}
%
In this section, 
we introduce $SU(3)$-${\bf 3 \times 3}$ order parameter and 
construct the effective potential 
referring to our previous paper \cite{NYS}.
%
\subsection{Structure of the order parameter}
%
The order parameter is given by the 3-by-3 matrix : 
$\Psi$$=$$(\Psi_{\alpha,\beta})$. 
For $g \in SU(3)$, it transforms as
\begin{equation}
     \Psi \longrightarrow g \Psi {}^tg, 
\label{eQaA}
\end{equation}
where ${}^tg$ is a transposition of $g$. 
It makes a ${\bf 3} \times {\bf 3}$-representation of $SU(3)$. 

Since permutation symmetry for indices $(\alpha, \beta)$ are conserved under the 
$SU(3)$-transformation, 
the order parameter $\Psi=(\Psi_{\alpha,\beta})$ can be decomposed into
the symmetric and Antisymmetric components: 
\begin{equation}
     S ={1 \over 2} (\Psi +{}^t\Psi),  \quad
     A ={1 \over 2} (\Psi -{}^t\Psi).   
\label{eQa}
\end{equation}
The $S$ and $A$ correspond to the {\bf 6} and ${\bf 3}^*$ representations 
which are irreducible representations :\, 
${\bf 3} \times {\bf 3}$$=$${\bf 3}^* + {\bf 6}$. 
From this consideration, 
symmetry breaking patterns for the condensed phases 
are classified into four types:
\begin{eqnarray}
\label{eQb}
     &\hbox{1) No-condensate}           &\quad A=S=0,     \label{con0}                 \\
     &\hbox{2) {\bf 3}${}^*$-condensate}&\quad A \neq 0, \quad S=0,  \label{con2}      \\
     &\hbox{3) {\bf 6}-condensate}      &\quad A=0,      \quad A \neq 0,  \label{con3} \\
     &\hbox{4) Mixed-condensate}        &\quad A \neq 0, \quad S \neq 0.  \label{con4}
\end{eqnarray}

Because of the $SU(3)$ invariance of the system, 
for $g \in SU(3)$, 
order parameters $(A,S)$ and $(g A {}^tg, g S {}^tg)$ 
represent the same condensed states, 
so that we can decompose the order parameters $A$ and $S$ as
\begin{equation}
     A =e^{i\theta} G A_0 {}^tG,  \quad
     S =e^{i\theta} G S_0 {}^tG,
\label{eQc}
\end{equation}
where $G \in SU(3)$ and $e^{i\theta} \in U(1)$ 
are the Goldstone degrees of freedom
(the system should also be invariant under the $U(1)$ transformation 
$\Psi \longrightarrow e^{i\theta} \Psi$).  
$A_0$ and $S_0$ in (\ref{eQc}) 
determine essentially scales of the symmetry breaking and 
can be parametrised as 
\begin{equation}
     A_0 = \pmatrix{    0   &  \phi_3 & -\phi_2 \cr
                    -\phi_3 &     0   &  \phi_1 \cr
                     \phi_2 & -\phi_1 &     0   \cr},  \quad
     S_0 = \pmatrix{ d_1  & 0   &  0  \cr
                     0    & d_2 &  0  \cr
                     0    & 0   & d_3 \cr}, 
\label{eQd}
\end{equation}
where $d_{1,2,3}$ are real and $\phi_{1,2,3}$ are complex parameters: 
$\phi_i =|\phi_i| e^{i\theta_i}$. 
%
\subsection{The Effective Potential for the order parameter}
%
The effective potential $V_{{\rm eff}}$ should be
a invariant function of $\Psi$  
under the $SU(3)$ and $U(1)$ transformations and 
constructed up to the fourth-order of $\Psi$ 
for a minimal Ginzburg-Landau model.
All the independent second-order and fourth-order terms in $\Psi$ 
are summed up: 
\begin{eqnarray}
     V^{(2)}_{{\rm eff}} &=& a_1 {\rm Tr} A^* A +a_2 {\rm Tr} S^* S 
\label{eQe} \\
     V_{{\rm eff}}^{(4)} 
         &=&b_1 ({\rm Tr} A^*A)^2 
         +b_2 ({\rm Tr} S^*S)^2 +b_3 {\rm Tr} S^*SS^*S \nonumber\\
         &+&b_4 {\rm Tr} A^*A {\rm Tr} S^*S 
           +b_5 {\rm Tr} A^*A S^*S 
           +b_+  ({\rm Tr} A^*SA^*S +\hbox{c.c}) 
           +ib_- ({\rm Tr} A^*SA^*S -\hbox{c.c}) \label{eQf}.
\end{eqnarray}
Inserting eq.~(\ref{eQc}) with (\ref{eQd}) into (\ref{eQe}, \ref{eQf}),  
the potentials $V^{(2,4)}_{{\rm eff}}$ can be represented by the variables
$R \equiv \sqrt{|\phi_1|^2 +|\phi_2|^2 +|\phi_3|^2}$ and 
$D \equiv \sqrt{d_1^2 +d_2^2 +d_3^2}$, 
and the normalised coordinates (angular parameters): 
$r_i \equiv |\phi_i|/R$ and $e_i \equiv d_i/D$: 
\begin{equation}
     V_{{\rm eff}}^{(2+4)} =g_R R^2 +g'_R R^4 
                           +g_D D^2 +(g'_D -h_D x) D^4
                           +(g''_D +h''_D u) R^2 D^2, 
\label{eQg}
\end{equation}
where the coefficients $g_i$ are represented 
as linear combinations of $a_i$ and $b_i$ in (\ref{eQe}, \ref{eQf}):
\begin{eqnarray}
     &&g_R =-2 a_1,  \quad
       g_R' =4 b_1,  \quad
       g_D =a_2,     \quad
       g_D'=b_2+b_3, \nonumber\\
     &&h_D =2 b_3,  \quad
       g''_D=-2 b_4 -b_5,  \quad
       h''_D=b_5. 
\label{eQi}
\end{eqnarray}
Variables $x$ and $u$ in (\ref{eQg}) are defined by 
\begin{equation}
     x \equiv e_1^2 e_2^2 +e_2^2 e_3^2 +e_3^2 e_1^2, 
\quad
     u \equiv f_1 r_1^2 +f_2 r_2^2 +f_3 r_3^2,     
\label{eQgA}
\end{equation}
where 
\begin{equation}
     f_1 =e_1^2 +c \sin(2\theta_1 +\delta) e_2 e_3,  \quad
     f_2 =e_2^2 +c \sin(2\theta_2 +\delta) e_3 e_1,  \quad
     f_3 =e_3^2 +c \sin(2\theta_3 +\delta) e_1 e_2. 
\label{eQgB}
\end{equation}
The constants $c$ and $\delta$ in (\ref{eQgB}) 
are defined with the coefficients $b_\pm$ in (\ref{eQf}) as 
$c =-4\sqrt{b_+^2+b_-^2}/b_5$ and $\tan{\delta} =b_+/b_-$. 

The range of variables $(u, x)$ is obtained in \cite{NYS}, 
which depends only on the parameter $c$ and 
is represented by the symbol, $\Lambda(c)$. 
%
\subsection{Stability condition}
%
For the infinitely large value of $R$ or $D$,
the effective potential should be positive definite.  
This stability (positive definite) condition gives some constraints 
on the parameters $g_i$ in (\ref{eQg}).
First of all, 
two constraints can be obtained 
from the behaviour of the effective potential (\ref{eQg}) 
when $D, R \to \infty$ : \ $g'_D \geq 0$ and $g'_R \geq 0$. 

We replace the variables $R$ and $D$ by the scaled ones: 
$R \to R/(g'_R)^{\frac{1}{4}}$ and $D \to D/(g'_D)^{\frac{1}{4}}$, 
then eq.~(\ref{eQg}) becomes
\begin{equation}
     V_{{\rm eff}}=-G R^2 +R^4 -F D^2 +(1+F' x) D^4 
                   +2 (H +H' u) R^2 D^2, 
\label{eQn}
\end{equation}
where scaled parameters are defined by
\begin{equation}
     G =-\frac{g_R}{\sqrt{g'_R}},  \quad
     F =-\frac{g_D}{\sqrt{g'_D}},  \quad
     F'=-\frac{h_D}{g'_D},         \quad
     H =\frac{g''_D}{2\sqrt{g'_D g'_R}},  \quad
     H'=\frac{h''_D}{2\sqrt{g'_D g'_R}}. 
\label{eQo}
\end{equation}
We use eq.~(\ref{eQn}) as a basic form of the effective potential in the following sections.

For large values of $R^2$ and $D^2$, 
eq.~(\ref{eQn}) behaves as a linear quadratic form of them
for any fixed values of $(x,u)$: 
$V_{{\rm eff}} \sim R^4 +(1+F' x) D^4 +2(H+H' u) R^2 D^2$. 
The positive definite condition for it is given by: 
\begin{eqnarray}
     &\hbox{a)}\ & 1+F' x -(H +H' u)^2 \ge 0 \nonumber \\
     && \quad \hbox{or}                    \nonumber \\
     &\hbox{b)}\ & 1 +F' x \ge 0 \quad\rm{and}\quad
                 H +H' u \ge 0. 
\label{eQq}
\end{eqnarray} 
In $(x,u)$-space the region satisfying inequalities (\ref{eQc}) is represented 
by the symbol, $\Sigma(F', H', H)$.  
This condition should be satisfied for 
any values of $(x,u)$ in $\Lambda(c)$, 
which geometrically means $\Lambda$$\subset$$\Sigma$ 
that puts some constraints on the coefficient parameters $c$, $F'$, $H$ and $H'$. 
These constraints (called ``the stability condition" below) 
were already obtained in \cite{NYS}.

From now on, 
we convert the variable ($x, u$) into ($U, X$), 
which are defined by : 
$X$$\equiv$$1+x F'$ and $U$$\equiv$$H+u H'$ 
for simplicity in calculations. 
Thus the region $\Lambda(c)$ is also redefined in ($U, X$)-space and 
represented by $\Lambda(F', H', H, c)$ 
(because of its dependence on $F'$, $H'$, $H$ and $c$).
$\Lambda(F', H', H, c)$ is shown as 
the area closed by lines or parabolic segments in Fig.~1 : 
\begin{itemize}
\item{ $0$$<$$c$$<$$2$ (Fig.~1-a)}
\begin{eqnarray}
 \overline{AE} &:& 
         X_1(U)=\frac{F'}{(cH')^2}(U-H)^2+1 \ \ 
         \left( H-\frac{cH'}{2}, H \right) \label{rmd1} \\
 \overline{AB} &:& 
         X_2(U, c) \ \ \quad \quad \quad 
         \left( H-\frac{cH'}{2}, H+\frac{1-c}{3} H' \right) \label{rmd2} \\
 \overline{CD} &:& 
         X_2(U, -c) \ \ \quad \quad \, \,  
         \left( H+\frac{1+c}{3} H', H+H' \right), \label{rmd3} 
\end{eqnarray}
where $X_2(U, c)$$\equiv$$-\frac{3F'}{(c+2)^2H'^2}(U-H-\frac{1-c}{3} H')^2+1+\frac{F'}{3}$ 
and we only give the parabolic segments $X_i(U)$ as functions of $U$ and 
the domains of $U$ in closing parentheses. 
 
\item{ $2$$<$$c$ (Fig.~1-b)}
\begin{eqnarray}
 \overline{AE}\  &:&
   X_1(U) \ \ \left( H-\frac{cH'}{2}, H \right) \  
             \hbox{and}\ 
\overline{GF} : X_1(U) \ 
            \left( H+\frac{cH'}{2}, U_c \right) \label{rmd4} \\
\overline{AB}\  &:&
  X_2(U, c) \ \ \left( H-\frac{cH'}{2}, H+\frac{(1-c)H'}{3} \right) \label{rmd5} \\
\overline{GC}\ &:&
  X_2(U, -c) \ \ \left( H+\frac{cH'}{2}, H+\frac{(c+1)H'}{3}  \right) \  
 \hbox{and}\ 
\overline{DF} : X_2(U, -c) \ 
\left( H+H', U_c \right), \label{rmd6}
\end{eqnarray}
where $U_c$$\equiv$$\frac{2 (c^2-c+1)H+c(2c-1)H'}{2 (c^2-c+1)}$ 
that is the intersection point of $X_1(U)$ and $X_2(U, -c)$.
\end{itemize}
It should be noticed that for each set of signs of $F'$ and $H'$,  
the corresponding $\Lambda(F', H', H, c)$ counterchanges each other by 
flip horizontal or vertical in the ($U, X$)-space.

The positive definite condition (\ref{eQq}) is also given in $(U, X)$-space:
\begin{eqnarray}
 X>U^2 \quad \hbox{or} \quad X>0 \ \hbox{and} \ U>0. \label{pdc}
\end{eqnarray}
The region satisfying inequalities (\ref{pdc}) have no dependence 
on the coefficient parameters so is represented by $\Sigma$.

Summarising this section, 
we give the positive definite condition $\Sigma$ and 
the possible range of ($U, X$) as $\Lambda(F', H', H, c)$.
And let us say again that stability condition,  
$\Lambda$$\subset$$\Sigma$, was already given in \cite{NYS} as the main results.
%
\section{A phase diagram}
%
In this section 
we derive a phase diagram of the effective potential eq.~(\ref{eQn}) 
which has four kinds of extrema ($R_0$, $D_0$) with respect to the variables $D$ and $R$ 
that correspond to the phases (\ref{con0})-(\ref{con4}) as shown in Table~1.
\begin{center} Table~1: Values of the order parameter and its potential. \\
     \begin{tabular}{| c | c | c | c |} \hline
                    &  $R_0^2$, $D_0^2$          & $V(R_0, D_0)$    & abbreviation  \\ \hline
   Normal-phase     &  $0, 0$                    & $V_0=0$          & {\bf N}-phase     \\ 
  {\bf $3^*$}-phase &  $\frac{G}{2}, 0$          & $V_3=-\frac{G^2}{4}$ 
                                                                    & {\bf A}-phase  \\ 
  {\bf $6$}-phase   &  $0, \frac{F}{2 X}$        & $V_6=-\frac{G^2}{4}\frac{Q^2}{X}$ 
                                                                    & {\bf S}-phase  \\
  Mixed-phase       &  $\frac{-FU+GX}{2 (X-U^2)}, 
                        \frac{F-GU}{2 (X-U^2)}$  & 
                 $V_m=-\frac{G^2}{4} \{ 1+\frac{(U-Q)^2}{X-U^2} \}$ & {\bf M}-phase \\ \hline
     \end{tabular}
\end{center}
In Table~1, 
we defined $Q$$\equiv$$\frac{F}{G}$ as a important parameter 
and for simplicity in calculations.  

In advance  
we show a procedure to derive the phase diagrams : 
\begin{description} 
\item{1)}\   
we firstly derive the region 
where each phase characterised by ($R_0^2$, $D_0^2$) in Table~1 is defined 
on the coefficients or in the ($U, X$)-space within the region $\Sigma$, (\ref{pdc}).
\item{2)}\ 
we secondly search the smallest one among $V_3$, $V_6$ and $V_m$ 
by directly comparing them within the region mentioned in 1)
where more than two kinds of phases are defined.
We first compare $V_6$ with $V_3$ and 
then $V_m$ with $V_6$ ($V_3$) that is smaller than $V_3$ ($V_6$).
Conclusively the region where $V_m$$<$$V_6$ ($V_3$) is obtained in the ($U, X$)-space.
We represent this region by the symbol, $\Gamma_{m6}$ ($\Gamma_{m3}$).
\item{3)}\ 
We thirdly put the constraints on the coefficients by enforcing 
the condition that in the ($U, X$)-space 
the region $\Gamma$ obtained in 2) contains a part of $\Lambda(F', H', H, c)$, 
$\Gamma$$\cap$$\Lambda$.
If this condition is satisfied the {\bf M}-phase can emerge as the most stable phase.
\item{4)}\ 
To obtain the complete phase diagram, 
we finally show the constraints on the coefficients obtained in 3) 
within the region where the stability condition is satisfied.  
\end{description} 
%
\subsection{the region where each phase is defined}
%
According to Table~1, 
the region where {\bf A}-phase is defined is easily obtained 
as the positive definite condition of ($R_0^2, D_0^2$):  
\begin{equation}
 0<G. \label{pmA}
\end{equation}
%

From the positive definite condition (\ref{pdc})  
one find that $X$$>$$0$.
Thus the possible region for {\bf S}-phase is obtained: 
\begin{equation}
 0<F \rightarrow 
                        \left\{ \begin{array}{l}
                                 0<G\ \hbox{and}\ 0<Q      \\
                                 G<0\ \hbox{and}\ Q<0
                                \end{array} \right.  \label{pmS}.
\end{equation}

According to Table~1, 
we shall obtain the possible region 
where {\bf M}-phase is defined in $(U, X)$ space: 
\begin{eqnarray}
 \hbox{If}\ \ G>0 \quad 
     &&\left\{ \begin{array}{l}
                U < Q, \ X > U^2 \ \ \hbox{and} \ \ X > QU \\
                U > Q, \ X < U^2 \ \ \hbox{and} \ \ X < QU
               \end{array} \right.
  \label{pm1} \\
 \hbox{If}\ \ G<0 \quad 
     &&\left\{ \begin{array}{l}
                U > Q, \ X > U^2 \ \ \hbox{and} \ \ X < QU \\
                U < Q, \ X < U^2 \ \ \hbox{and} \ \ X > QU
               \end{array} \right. . 
 \label{pm2} 
\end{eqnarray}
These conditions (\ref{pm1}-\ref{pm2}) are shown in Fig.~2 for fixed ($G$, $Q$)
within the positive definite condition (\ref{pdc}).
%
\subsection{comparison $V_3$ with $V_6$}
%
According to the above subsection the possible region where
both {\bf A}-phase and {\bf S}-phase are defined is ($0$$<$$G$, $0$$<$$Q$).
Because of the definition of $X$$\equiv$$1$$+$$F'x$ and 
the possible range of $x$ : $0$$\le$$x$$\le$$\frac{1}{3}$ \cite{NYS}, 
the minimum value of $V_6$$=$$-\frac{G^2}{4}\frac{Q^2}{X}$ 
has two kinds by the signs of $F'$: \ 
$V_6$$=$$-\frac{G^2}{4}Q^2$ ($F'$$>$$0$) and 
$V_6$$=$$-\frac{G^2}{4}\frac{Q^2}{1+\frac{F'}{3}}$ ($F'$$<$$0$). 
So we obtain the condition that $V_6$$<$$V_3$ for ($0$$<$$G$, $0$$<$$Q$): 
\begin{eqnarray}
   \hbox{If}\ \ 1 < Q, && \quad -3 < F'                       \nonumber \\
   \hbox{If}\ \ 0< Q \leq 1, && \quad -3 < F' <  -3(1-Q^2).   \label{v361}
\end{eqnarray}
%
\subsection{comparison $V_m$ with $V_6$}
%
%
In the case that only {\bf S}- and {\bf M}-phase are defined ($G$$<$$0$, $Q$$<$$0$)
or (\ref{v361}) is satisfied, 
we should compare $V_m$ with $V_6$. 

By employing $V_6$$=$$-\frac{G^2}{4}Q^2$ (if $F'$$>$$0$) and 
$V_6$$=$$-\frac{G^2}{4}\frac{Q^2}{1+\frac{F'}{3}}$ (if $F'$$<$$0$) and 
comparing them with $V_m$ in Table~1, 
the regions where $V_m$$<$$V_6$ are obtained: 
\begin{eqnarray}
 \hbox{For}\ 1<|Q| \ \ \ 
        &\hbox{If}&\ -3<F'<0 \quad  
\left\{ \begin{array}{l}
          X > U^2 \quad \hbox{and} \quad X < X_4(U, F') \\
          X < U^2 \quad \hbox{and} \quad X > X_4(U, F')
        \end{array} \right. 
          \label{v6m1} \\
        &\hbox{If}&\  0<F' \quad \quad \quad \ \ 
\left\{ \begin{array}{l}
          X > U^2 \quad \hbox{and} \quad X < X_4(U, 0) \\
          X < U^2 \quad \hbox{and} \quad X > X_4(U, 0)
        \end{array} \right.
        \label{v6m2} \\
{} \nonumber \\
\hbox{For}\  |Q|<1 \ \ \ 
       &\hbox{If}&\ -3< F'<-3(1-Q^2) \quad
\left\{ \begin{array}{l}
          X > U^2 \quad \hbox{and} \quad X < X_4(U, F') \\
          X < U^2 \quad \hbox{and} \quad X > X_4(U, F')
        \end{array} \right. 
        \label{v6m3} \\
        &\hbox{If}&\  -3(1-Q^2)< F'<0 \quad \ \, \, 
\left\{ \begin{array}{l}
          X > U^2 \quad \hbox{and} \quad X > X_4(U, F') \\
          X < U^2 \quad \hbox{and} \quad X < X_4(U, F')
        \end{array} \right. 
        \label{v6m4} \\
        &\hbox{If}&\  0<F' \quad 
         \quad \quad \quad \quad \quad \quad \ \ \ 
\left\{ \begin{array}{l}
          X > U^2 \quad \hbox{and} \quad X > X_4(U, 0) \\
          X < U^2 \quad \hbox{and} \quad X < X_4(U, 0)
        \end{array} \right. , 
        \label{v6m5} 
\end{eqnarray}
where
\begin{equation} 
X_4(U, F') \equiv \frac{Q^2}{Q^2-(1+\frac{F'}{3})}
         \left( U-\frac{1+\frac{F'}{3}}{Q} \right)^2+1+\frac{F'}{3}.
\end{equation}

These inequalities do not hold the condition 
in which {\bf M}-phase is defined in the $(U, X)$-space.
So we should further
investigate the region where one fined above regions (\ref{v6m1}-\ref{v6m5})
within the region defined by (\ref{pdc}, \ref{pm1}-\ref{pm2}) (Fig.~2), 
which is represented as $\Sigma_{m6}(G, Q, F')$. 
Combining the conditions (\ref{pdc}, \ref{pm1}-\ref{pm2})
with (\ref{v6m1}-\ref{v6m5}),  
we obtained $\Gamma_{m6}(F, G, F')$ and show it in Fig.~3 for ($0$$<$$G$, $0$$<$$Q$) 
and Fig.~4 for ($G$$<$$0$, $Q$$<$$0$).
%
\subsection{comparison $V_m$ with $V_3$}
%
In the case that the condition (\ref{v361}) is not satisfied,  
that means $V_3$$<$$V_6$, 
or only {\bf M}-phase and {\bf S}-phase are defined ($0$$<$$G$, $Q$$<$$0$), 
we shall compare $V_m$ with $V_3$.
In a similar way of the above case $V_m$$<$$V_6$, 
we obtained the region $\Gamma_{m3}(G, Q, F')$ where $V_m$$<$$V_3$
as shown in Fig.~5.
%
\subsection{Derivation of a phase diagram}
%
In the above sections 
we obtained the regions where $V_m$ is the smallest one in the ($U, X$)-space,    
which is represented by $\Gamma_{m6}$ and $\Gamma_{m3}$. 
But if $\Gamma$ is outside of the range of ($U, X$), 
{\bf M}-phase can not emerge as the most stable phase.  

So if the range $\Lambda(F', H', H, c)$ 
includes a part of $\Gamma_{m6}(F, G, F')$ 
{\bf M}-phase emerges : 
\begin{equation}
 \Gamma_{m6}(G, Q, F') \cap \Lambda(F', H', H, c). \label{con1}
\end{equation}
If not, {\bf S}-phase can emerge 
because the condition (\ref{v361}) 
was already satisfied.

This situation is available in the case of {\bf A}-phase and {\bf M}-phase : 
\begin{equation}
 \Gamma_{m3}(G, Q, F') \cap \Lambda(F', H', H, c). \label{con2}
\end{equation}

The conditions (\ref{con1}-\ref{con2}) put some constraints 
on the coefficients ($G$, $Q$, $F'$, $H'$, $H$ and $c$). 
According to these constraints we can obtain the phase diagrams as
the main results of this paper.

Since there are six independent parameters ($H$, $F'$, $c$, $H'$, $G$ and $Q$), 
one can notice that for a fixed set of ($H$, $F'$), 
if the stability condition is given for a set ($c$, $H'$),  
the constraints from (\ref{con1}-\ref{con2})
can be classified by ($G$, $Q$).
So we represent a phase diagram in ($H$, $F'$) space 
(where the stability condition is already given for a fixed set of ($c$, $H'$))
for a fixed set of ($G$, $Q$) relying on the above results.

As an example, 
let's consider the case that ($0$$<$$c$$<$$2$, $0$$<$$H'$) and
($0$$<$$G$, $0$$<$$Q$).
\begin{itemize}
 \item If $Q>1$  \\
In this case, $V_3$$>$$V_6$ in the whole region of $-3$$<$$F'$ from (\ref{v361}).

For $0$$<$$F'$, 
$\Gamma_{m6}$ and $\Lambda(F', H', H, c)$ are illustrated in Fig.~6. 
The condition (\ref{con1}) is satisfied if
the segment $\overline{AB}$ of $\Lambda$ intersects 
$\overline{CD}$ of $\Gamma_{m6}$:  
\begin{eqnarray}
 && H < \frac{1}{Q}  \label{pha1} \\
 \hbox{or} \ \ 
 && \frac{1}{Q} < H < \frac{1}{Q}+\frac{cH'}{2} \ \ \hbox{and}\ \ 
    F' < F'_3(H, c),  \label{pha2}
\end{eqnarray}
where $F'_3(H, c)$$\equiv$$\frac{(2QH-2-cQH')^2}{Q^2-1}$.

For $-3$$<$$F'$$<$$0$, 
similarly to the case $0$$<$$F'$,  
the condition (\ref{con1}) depends only on 
the boundaries of $\Gamma_{m3}$ and $\Lambda$: 
\begin{eqnarray}
 && F' > 3QH-3-(c-1)QH' \label{pha3} \\
 \hbox{or}\ \ 
 &&  F' < 3QH-3-(c-1)QH'\ \hbox{and}\ 
     F' > F'_{+}(H,c) \label{pha4}
\end{eqnarray}
where {\footnotesize $F'_{\pm}(H, c)$$\equiv$$\frac{1}{2}
\left( 8QH-4cQH'-Q^2-7 \pm \sqrt{-12(2QH-cQH'-2)^2+(8QH-4cQH'-Q^2-7)^2}\right)$ }.
According to eqs.(\ref{pha1}-\ref{pha4}), 
the phase diagram in $(H, F')$ space for the set ($0<c<2$, $1<Q$, $0<G$ and $0<H'$)
is obtained especially with stability condition 
for $0$$<$$H'$$<$$\frac{3}{c+2}$ \cite{NYS} in Fig.~9-f.
 \item If $0<Q<1$ \\
In this case, 
it is seen from eq.~(\ref{v361}) that 
$V_6$$>$$V_3$ if $F'$$>$$-3(1-Q^2)$ and
$V_3$$>$$V_6$ if $-3$$<$$F'$$<$$-3(1-Q^2)$.

Similarly to the case $1$$<$$Q$, 
we obtain the conditions (\ref{con2}) and (\ref{con1}): 
\begin{eqnarray}
 &&F' > -3(1-Q^2) \ \ \hbox{and}\ \ H < \frac{cH'}{2}+Q \label{pha5} \\
 &&-3 < F' < -3(1-Q^2)  \ \ \hbox{and}\ \ 
     \left\{ \begin{array}{l}
                 F' > 3QH-3-(c-1)QH'  \\
   \hbox{or} \ \ F' < 3QH-3-(c-1)QH' \ \hbox{and}\ F' > F'_+(H, c)
             \end{array} \right. . \label{pha6}
\end{eqnarray}
\end{itemize}
These are summarised and shown in Fig.~9-e.

Other cases are also studied in the same fashion, 
the complete phase diagram is organised in the next section.
%
\section{Results and Conclusions}
%
In this section 
we organise the resulting phase diagrams in the ($H$, $F'$)-space in the systematic way.
We firstly classify them by 
two divisions of $c$ : $0$$<$$c$$<$$2$ and $2$$<$$c$ \, 
( these two come from the possible range of ($x, u$) 
that relates to the angular parameter of the order parameter ), \, 
secondly by 
four divisions of $H'$ : $H'$$<$$-\frac{3}{|2-c|}$, $-\frac{3}{|2-c|}$$<$$H'$$<$$0$, 
                      $0$$<$$H'$$<$$\frac{3}{2+c}$, $\frac{3}{2+c}$$<$$H'$ \, 
(they are used to classify the stability condition) 
and finally by the sets of ($G, Q$) 
that are the coefficients of the second-order terms in the effective potential (\ref{eQn}) 
via $Q\equiv\frac{F}{G}$. 
It is helpful to attend the signs of $F$ and $G$, that is,  
if $G$$>$$0$ ($F$$>$$0$) the {\bf A}({\bf S})-phase has a potential to emerge.

Ahead of listing the phase diagrams 
we give the functions to appear in the tables below and in corresponding figures : 
\begin{eqnarray}
&&H_1(F', c) \equiv \frac{1}{Q}+\frac{cH'}{2}+\frac{2F'-\sqrt{F'(3-3Q^2+F'}}{6Q} \\
&&F'_{\pm}(H, c) \equiv \frac{1}{2}
\left( 8QH-4cQH'-Q^2-7 \right.  \nonumber \\ 
 && \quad \quad \quad \quad \quad 
   \left. \pm \sqrt{-12(2QH-cQH'-2)^2+(8QH-4cQH'-Q^2-7)^2}\right) \\
&&F'_1(H) \equiv \frac{-3(1-QH-QH')}{1-2QH-2QH'+Q^2} \\
&&F'_2(H, c) \equiv 3QH+(c+1)QH'-3  \\
&&F'_3(H, c) \equiv \frac{(2QH-2-cQH')^2}{Q^2-1}, 
\end{eqnarray}
where $H_1(F', c)$ is the inverse function of $F'_{\pm}(H, c)$.
The constants : 
\begin{eqnarray}
&&H_a(c) \equiv \frac{1}{6Q}\left\{ 3(Q^2+1)-2QH'(c+1)
               -\sqrt{9(Q^2-1)^2+4(c-2)^2H'^2Q^2}  \right\} \\
&&F'_a(c) \equiv \frac{1}{2}\left\{ 3(Q^2-1)-\sqrt{9(Q^2-1)^2+4(c-2)^2H'^2Q^2}  \right\} \\
&&F'_b(c) \equiv \frac{1}{2}\left[ 3(Q^2-1)+2(c-2)H'
                -\sqrt{ (Q^2-1)\left\{ \left( 2(c-2)H'+3Q \right)^2-9 \right\} }  \right].
\end{eqnarray}
We also define the functions used for the stability conditions :
\begin{eqnarray}
&&F'_I(H, c)\equiv \frac{3}{2} \left[ 
       \left( H +H' \right) \left( H +\frac{2c -1}{3}H' \right) -1 \right.  \nonumber \\
 && \quad \quad \quad \quad \quad \quad \quad 
               \left. - \sqrt{\left\{ (H +H')^2 -1 \right\} 
                          \left\{ \left( H +\frac{2c -1}{3}H' \right)^2-1 \right\}}  
                                                                      \right] \\
&&F'_{II}(H, c)\equiv (2H-cH')^2-4 \\
&&F'_{III}(H)\equiv \frac{(cH')^2}{1-H^2}.
\end{eqnarray}
%
%

In the following tables (Table~2-Table~7), 
we present the inequalities that 
just indicate the most stable phases to be inside the stability condition 
that has a role as the border line of the stable phases.
But in corresponding figures, 
we show the complete phase diagrams with the stability condition 
(through all the figures of phase diagrams, 
$F'$-axis is vertical and $H$-axis is horizontal).

In each table, 
we show the numbered entry in the first column,  
the sets of ($G, Q$) in the second one, 
the conditions for each phase emerging in the ($H, F'$)-space 
(in the closing parentheses the domain of $H$ for a function $F'(H)$ is given) 
in the third one 
and the figure numbers of the phase diagrams in the fourth one.
\newpage
\begin{center}
Table~2 : The phase diagrams 
for $0$$<$$c$$<$$2$ and $H'$$<$$-\frac{3}{2-c}$.
\end{center}
\begin{center}
\begin{tabular}{lllc}              \hline \hline
\multicolumn{4}{c}{ \textbf{$H'$$<$$-\frac{3}{2-c}$} }      \\ \hline
\multicolumn{1}{c}{No. }   &
\multicolumn{1}{c}{$G$, $Q$}   & 
\multicolumn{1}{c}{conditions} & 
\multicolumn{1}{c}{figures}  \\ \hline
1-1\ &  $G$$<$$0$, $Q$$<$$-1$ & \quad {\bf M}-phase : $F'>F'_1(H)$ \quad  
                                         $\left( -H'-1, -H'+\frac{1}{Q} \right)$ & Fig.~7-a  \\
     &                        & \quad {\bf S}-phase : \hbox{remaining region}    & \\
1-2\ & $G$$<$$0$, $-1$$<$$Q$$<$$0$ 
                              & \quad {\bf S}-phase : whole region & \\ 
1-3\ & $G$$<$$0$, $0$$<$$Q$   & \quad {\bf N}-phase : whole region & \\
1-4\ & $0$$<$$G$, $Q$$<$$0$   & \quad {\bf M}-phase : $H<-H'+Q$, \,     
                                      If $Q$$<$$-1$ {\bf M}-phase vanishes. & Fig.~7-b \\ 
     &                        & \quad {\bf A}-phase : remaining region.     & \\
1-5\ & $0$$<$$G$, $0$$<$$Q$$<$$1$ &
          \quad {\bf M}-phase : $F'>F'_1(H)$ \quad $\left( -H'-1, -H'+Q \right)$ & Fig.~7-c \\
     &   &\quad {\bf A}-phase : $-H'+Q<H$ and $F'>-3(1-Q^2)$ & \\
     &   &\quad {\bf S}-phase : remaining region & \\
1-6\ & $0$$<$$G$, $1$$<$$Q$  &\quad the same as 1-1   & \\ \hline 
\end{tabular}
\end{center}
\newpage
\begin{center}
Table~3 : The phase diagrams 
for $0$$<$$c$$<$$2$ and $-\frac{3}{2-c}$$<$$H'$$<$$0$.
\end{center}
\begin{center}
\begin{tabular}{lllc}               \hline \hline
\multicolumn{4}{c}{ \textbf{$-\frac{3}{2-c}$$<$$H'$$<$$0$}}      \\ \hline
\multicolumn{1}{c}{No. }   &
\multicolumn{1}{c}{$G$, $Q$}   & 
\multicolumn{1}{c}{conditions} & 
\multicolumn{1}{c}{figures}  \\ \hline
2-1\ &  $G$$<$$0$, $Q$$<$$-1$ & \quad {\bf M}-phase :  $H<H_a(c)$,              & Fig.~8-a \\ 
     &                & \quad \quad \quad \quad \quad \quad $F'>F'_1(H)$ \ \  
                                $\left( H_a(c), -H'+\frac{1}{Q} \right)$,               & \\
     &                        & \quad \quad \quad \quad \quad \quad $F'<F'_2(H, c)$ \ \   
                                $\left( H_a(c), -\frac{(c+1)H'}{3} \right)$             & \\
     &                        & \quad {\bf S}-phase : \hbox{remaining region}           & \\
2-2\ & $G$$<$$0$, $-1$$<$$Q$$<$$0$ & \quad {\bf M}-phase : 
                                   $F'<F'_2(H, c)$ \ \  
                                   $\left( H_a(c), -\frac{(c+1)H'}{3} \right)$ &  Fig.~8-b \\   
     &                             & \quad {\bf S}-phase : remaining region             & \\ 
2-3\ & $G$$<$$0$, $0$$<$$Q$   & \quad {\bf N}-phase : whole region                      & \\
2-4\ & $0$$<$$G$, $Q$$<$$0$   & \quad {\bf M}-phase : $H<Q-H'$, \,     
                                      If $Q$$<$$-1$ {\bf M}-phase vanishes.  & Fig.~8-c \\ 
     &                        & \quad {\bf A}-phase : remaining region.                 & \\
2-5\ & $0$$<$$G$, $0$$<$$Q$$<$$1$ &
              \quad {\bf M}-phase : $H<-\frac{(c+1)H'}{3}$,                  & Fig.~8-d \\
     &                            & \quad \quad \quad \quad \quad \quad
                                $F'>F'_1(H)$ \ \ $\left( H_a(c), Q-H' \right)$,        & \\
     &        & \quad \quad \quad \quad \quad \quad $F'>F'_2(H, c)$ \ \   
                                $\left(  -\frac{(c+1)H'}{3}, H_a(c) \right)$            & \\
     &       &\quad {\bf A}-phase : $Q-H'<H$ and $F'>-3(1-Q^2)$                        & \\
     &       &\quad {\bf S}-phase : remaining region & \\
2-6\ & $0$$<$$G$, $1$$<$$Q$  &\quad the same as 2-1 
               but $F'<F'_2(H, c)$ $\rightarrow$ $F'_2(H, c)<F'$ &  Fig.~8-e \\ \hline \hline
\end{tabular}
\end{center}
\newpage
\begin{center}
Table~4 : The phase diagrams 
for $0$$<$$c$$<$$2$ and $0$$<$$H'$$<$$\frac{3}{2+c}$.
\end{center}
\begin{center}
\begin{tabular}{lllc}               \hline \hline
\multicolumn{4}{c}{ \textbf{$0$$<$$H'$$<$$\frac{3}{2+c}$}}      \\ \hline
\multicolumn{1}{c}{No. }   &
\multicolumn{1}{c}{$G$, $Q$}   & 
\multicolumn{1}{c}{conditions} & 
\multicolumn{1}{c}{figures}  \\ \hline
3-1\ &  $G$$<$$0$, $Q$$<$$-1$ & \quad {\bf M}-phase : 
                                $0<F'$ and $H<\frac{1}{Q}$,         & Fig.~9-a, b  \\ 
     &                        & \quad \quad \quad \quad \quad \quad $0<F'<F'_3(H, c)$ \ \  
                                $\left( \frac{1}{Q}, \frac{cH'}{2}+\frac{1}{Q} \right)$,@& \\
     &                        & \quad \quad \quad \quad \quad \quad 
                               $F'<0$ and $F'<F'_2(H, -c)$,   & \\   
     &                        & \quad \quad \quad \quad \quad \quad 
                                       $F'_2(H, -c)<F'<0$ and $H<H_1(F', c)$            & \\
     &                        & \quad {\bf S}-phase : \hbox{remaining region}           & \\
3-2\ & $G$$<$$0$, $-1$$<$$Q$$<$$0$ & \quad {\bf M}-phase : 
                                     $F'<-3(1-Q^2)$ and              & Fig.~9-c, d  \\
     &              & \quad \quad \quad \quad \quad \quad 
                                    $H<H_1(F', c)$ and $F'_2(H, -c)<F'$,                & \\
     &                        & \quad \quad \quad \quad \quad \quad 
                                    $F'<-3(1-Q^2)$ and $F'<F'_2(H, -c)$     & \\   
     &                        & \quad {\bf S}-phase : remaining region                  & \\ 
3-3\ & $G$$<$$0$, $0$$<$$Q$   & \quad {\bf N}-phase : whole region                      & \\
3-4\ & $0$$<$$G$, $Q$$<$$0$   & \quad {\bf M}-phase : $H$$<$$\frac{cH'}{2}$$+$$Q$, \,      
                             If $Q$$<$$-\frac{2+cH'}{2}$ {\bf M}-phase vanishes.        & \\ 
     &                        & \quad {\bf A}-phase : remaining region.                 & \\
3-5\ & $0$$<$$G$, $0$$<$$Q$$<$$1$ &
              \quad {\bf M}-phase :  $H<\frac{(c-1)H'}{3}$,              & Fig.~9-e  \\
     &              & \quad \quad \quad \quad \quad \quad $F'>F'_2(H, -c)$ \ \ 
                     $\left( \frac{(c-1)H'}{3}, H_a(-c) \right)$,                   & \\
     &              & \quad \quad \quad \quad \quad \quad $F'>F'_{+}(H, c)$ \ \ 
                     $\left( H_a(-c), \frac{cH'}{2}+Q \right)$,                      & \\
     &       &\quad {\bf A}-phase : $\frac{cH'}{2}+Q<H$ and $F'>-3(1-Q^2)$            & \\
     &       &\quad {\bf S}-phase : remaining region & \\
3-6\ & $0$$<$$G$, $1$$<$$Q$  &\quad {\bf M}-phase : for $0<F'$ same as 3-1,   & Fig.~9-f  \\
     &                       & \quad \quad \quad \quad \quad \quad 
                                for $F'<0$, $F'_2(H, -c)<F'$,         & \\   
     &                       & \quad \quad \quad \quad \quad \quad 
                               $F'<F'_2(H, -c)$ and $H<H_1(F', c)$     & \\ 
     &       &\quad {\bf S}-phase : remaining region & \\\hline
\end{tabular}
\end{center}
\newpage
\begin{center}
Table~5 : The phase diagrams 
for $0$$<$$c$$<$$2$ and $\frac{3}{2+c}$$<$$H'$.
\end{center}
\begin{center}
\begin{tabular}{lllc}               \hline \hline
\multicolumn{4}{c}{\textbf{$\frac{3}{2+c}$$<$$H'$}}      \\ \hline
\multicolumn{1}{c}{No. }   &
\multicolumn{1}{c}{$G$, $Q$}   & 
\multicolumn{1}{c}{conditions} & 
\multicolumn{1}{c}{figures}  \\ \hline
4-1\ &  $G$$<$$0$, $Q$$<$$-1$ & \quad {\bf M}-phase : 
                                for $0<F'$ same as 3-1 &  Fig.~10-a, b \\
     &                        & \quad \quad \quad \quad \quad \quad 
                                for $F'<0$, $H<H_1(F', c)$                              & \\
     &                        & \quad {\bf S}-phase : \hbox{remaining region}           & \\
4-2\ & $G$$<$$0$, $-1$$<$$Q$$<$$0$ & \quad {\bf M}-phase : 
                                     $F'<-3(1-Q^2)$ and $H<H_1(F', c)$    & Fig.~10-c, d  \\  
     &                        & \quad {\bf S}-phase : remaining region                  & \\ 
4-3\ & $G$$<$$0$, $0$$<$$Q$   & \quad {\bf N}-phase : whole region                      & \\
4-4\ & $0$$<$$G$, $Q$$<$$0$   & \quad the same as 3-4            & \\
4-5\ & $0$$<$$G$, $0$$<$$Q$$<$$1$ &
              \quad {\bf M}-phase : $H<\frac{cH'-1}{2}$,                     &  Fig.~10-e \\
     &              & \quad \quad \quad \quad \quad \quad $F'>F'_{+}(H, c)$ \ \ 
                     $\left( \frac{cH'-1}{2}, \frac{cH'}{2}+Q \right)$,                  & \\
     &       &\quad {\bf A}-phase : $\frac{cH'}{2}+Q<H$ and $F'>-3(1-Q^2)$                        & \\
     &       &\quad {\bf S}-phase : remaining region                                     & \\
4-6\ & $0$$<$$G$, $1$$<$$Q$  &\quad the same as 4-1                &  Fig.~10-f \\  \hline
\end{tabular}
\end{center}
%
%
\newpage
\begin{center}
Table~6 : The phase diagrams 
for $2$$<$$c$ and $H'$$<$$-\frac{3}{c-2}$.
\end{center}
\begin{center}
\begin{tabular}{lllc}               \hline \hline
\multicolumn{4}{c}{\textbf{$H'$$<$$-\frac{3}{c-2}$}}      \\ \hline
\multicolumn{1}{c}{No. }   &
\multicolumn{1}{c}{$G$, $Q$}   & 
\multicolumn{1}{c}{conditions} & 
\multicolumn{1}{c}{figures}  \\ \hline
5-1\ &  $G$$<$$0$, $Q$$<$$-1$ & \quad {\bf M}-phase : $H<\frac{1}{Q}-H'$,   & Fig.~11-a, b  \\   
     &                 & \quad \quad \quad \quad \quad \quad $0<F'<F'_3(H, -c)$ \ \
                             $\left( \frac{1}{Q}-H', Q-\frac{cH'}{2} \right)$,          & \\
     &                 & \quad \quad \quad \quad \quad \quad 
                             $F'<0$ and $H<H_1(F', -c)$                                 & \\
     &                        & \quad {\bf S}-phase : \hbox{remaining region}           & \\
5-2\ & $G$$<$$0$, $-1$$<$$Q$$<$$0$ & \quad the same as 4-2 but $c$ $\rightarrow$ $-c$   & \\ 
5-3\ & $G$$<$$0$, $0$$<$$Q$   & \quad {\bf N}-phase : whole region                      & \\
5-4\ & $0$$<$$G$, $Q$$<$$0$   & \quad the same as 3-4 but $c$ $\rightarrow$ $-c$        & \\
5-5\ & $0$$<$$G$, $0$$<$$Q$$<$$1$ & \quad the same as 4-5 but $c$ $\rightarrow$ $-c$    & \\
5-6\ & $0$$<$$G$, $1$$<$$Q$  &\quad the same as 5-1                            & \\  \hline
\end{tabular}
\end{center}
\newpage
\begin{center}
Table~7 : The phase diagrams 
for $2$$<$$c$ and $-\frac{3}{c-2}$$<$$H'$$<$$0$.
\end{center}
\begin{center}
\begin{tabular}{lllc}               \hline \hline
\multicolumn{4}{c}{\textbf{$-\frac{3}{c-2}$$<$$H'$$<$$0$}}      \\ \hline
\multicolumn{1}{c}{No. }   &
\multicolumn{1}{c}{$G$, $Q$}   & 
\multicolumn{1}{c}{conditions} & 
\multicolumn{1}{c}{figures}  \\ \hline
6-1\ &  $G$$<$$0$, $Q$$<$$-1$ & \quad {\bf M}-phase : $H<\frac{1}{Q}-H'$, & Fig.~11-c, d  \\   
     &                 & \quad \quad \quad \quad \quad \quad $0<F'<F'_3(H, -c)$ \ \
                             $\left( \frac{1}{Q}-H', Q-\frac{cH'}{2} \right)$,          & \\
     &                 & \quad \quad \quad \quad \quad \quad 
                                 $F'<0$ and $F'<F'_2(H, c)$,                            & \\
     &                 & \quad \quad \quad \quad \quad \quad 
                             $F'_2(H, c)<F'<0$ and $H<H_1(F', -c)$                      & \\
     &                        & \quad {\bf S}-phase : remaining region                  & \\
6-2\ & $G$$<$$0$, $-1$$<$$Q$$<$$0$ & \quad the same as 3-2 but $c$ $\rightarrow$ $-c$   & \\ 
6-3\ & $G$$<$$0$, $0$$<$$Q$   & \quad {\bf N}-phase : whole region                      & \\
6-4\ & $0$$<$$G$, $Q$$<$$0$   & \quad the same as 5-4                                   & \\
6-5\ & $0$$<$$G$, $0$$<$$Q$$<$$1$ & \quad the same as 3-5 but $c$ $\rightarrow$ $-c$    & \\
6-6\ & $0$$<$$G$, $1$$<$$Q$  & \quad for $0<F'$ the same as 5-6,                        & \\
     &                       & \quad 
                             for $F'<0$ the same as 3-6 but $c$ $\rightarrow$ $-c$   & \\
 \hline
\end{tabular}
\end{center}

For the division $2<c$ and $0<H'$, 
the phase diagrams coincide completely 
with ones for $0<H'$ of $0<c<2$.

In the above tables and figures in them, 
we give a complete phase diagram of the effective potential (\ref{eQn}). 
We also expect that this model is useful to 
describe the dynamics of the condensed phase, e.g. 
vortex or the other topological excitations, 
as future works.
%
%
%
%

%
\newpage
\begin{description}
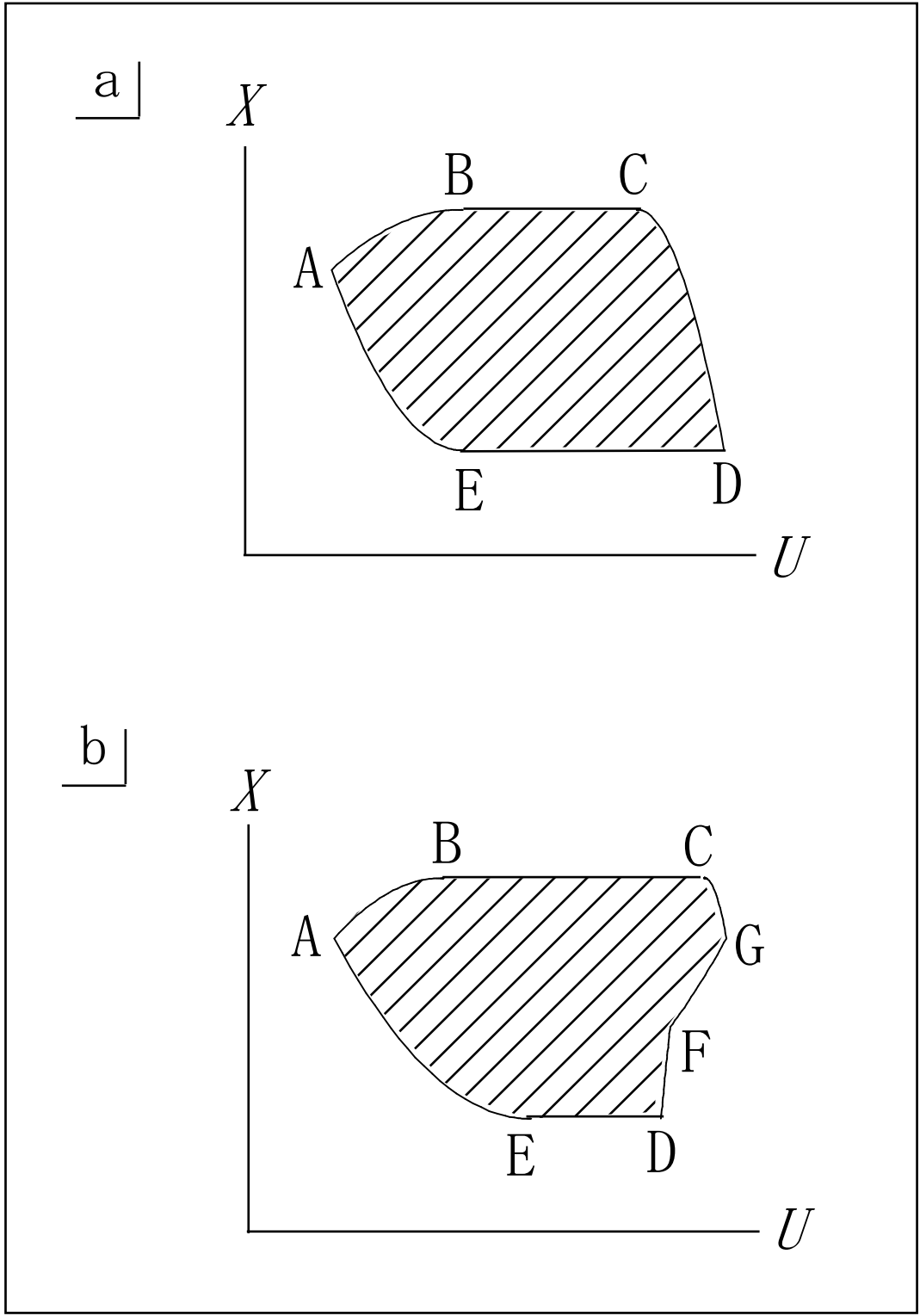
\item{Fig.~1.} 
The possible range of ($U$, $X$) 
specially for $0$$<$$F'$ and $0$$<$$H'$.
a) $0<c<2$, \,  
b) $2<c$. 
\item{Fig.~2.}
The region where {\bf M}-phase is defined, 
which is represented as crosshatched region and 
classified by $G$ and $Q$. 
a) : $0$$<$$G$, $0$$<$$Q$, 
b) : $0$$<$$G$, $Q$$<$$0$, 
c) : $G$$<$$0$, $Q$$<$$0$.
The common point $C$ is ($Q$, $Q^2$).
$\overline{OA}$ is $X$$=$$U^2$.
$\overline{OB}$ is $X$$=$$QU$.
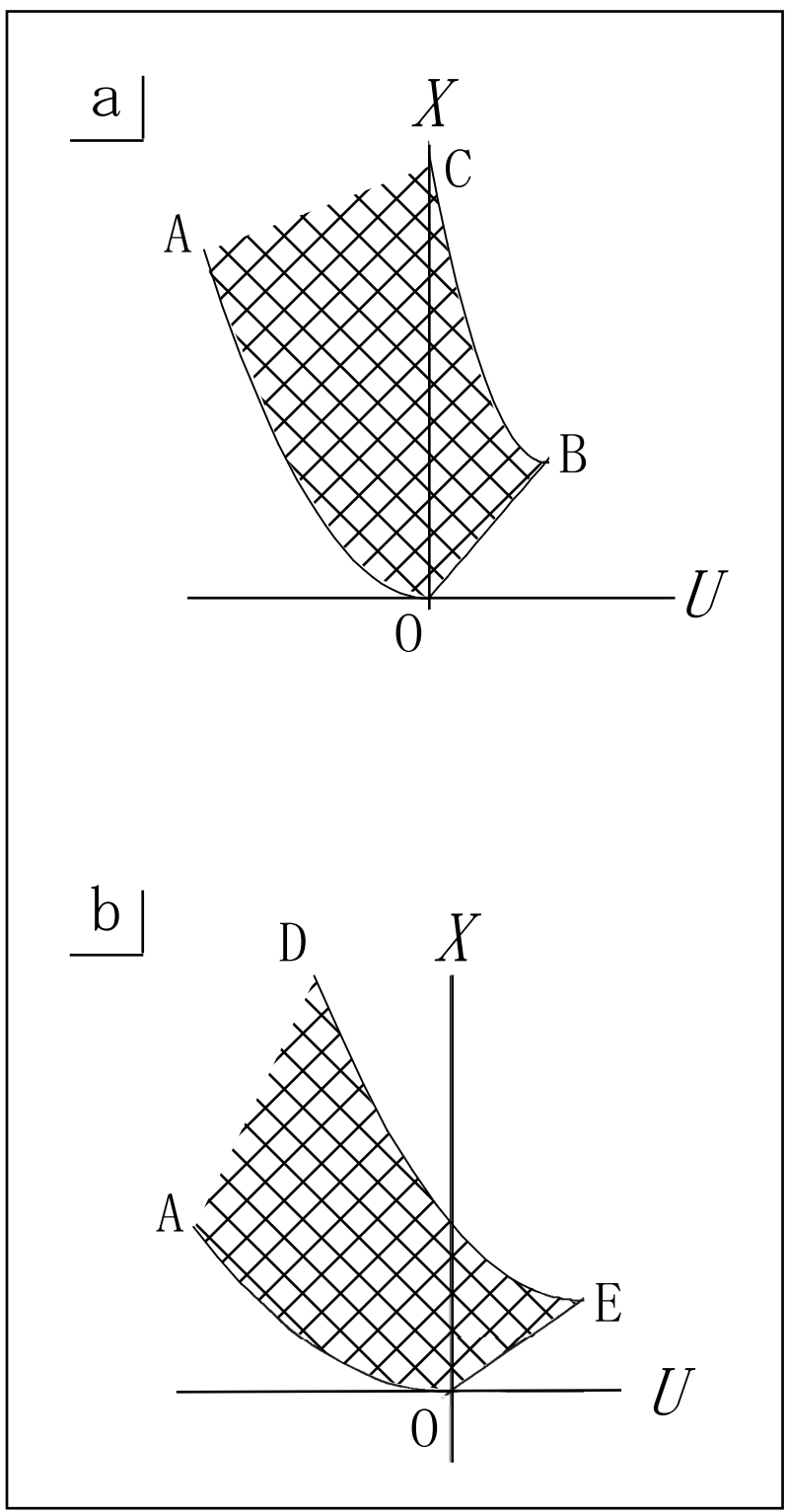
\item{Fig.~3.}
The region where both of {\bf M}- and {\bf S}-phase are defined 
and also $V_m$$<$$V_6$ is satisfied for ($0$$<$$G$, $0$$<$$Q$). 
These are classified by $Q$ and $F'$. \ 
a) : $1$$<$$Q$, $0$$<$$F'$. \ 
     $\overline{OA}$ is $X$$=$$U^2$.  
     $\overline{BC}$ is $X$$=$$X_4(U,0)$.
     The point $B$ is $\left( \frac{1}{Q}, 1 \right)$.
b) : ($1$$<$$Q$, $-3$$<$$F'$$<$$0$) or 
     ($0$$<$$Q$$<$$1$, $-3$$<$$F'$$<$$-3(1-Q^2)$). \ 
     $\overline{DE}$ is $X$$=$$X_4(U,F')$.
     The point $D$ is $\left( \frac{1+F'/3}{Q}, 1+F'/3 \right)$.
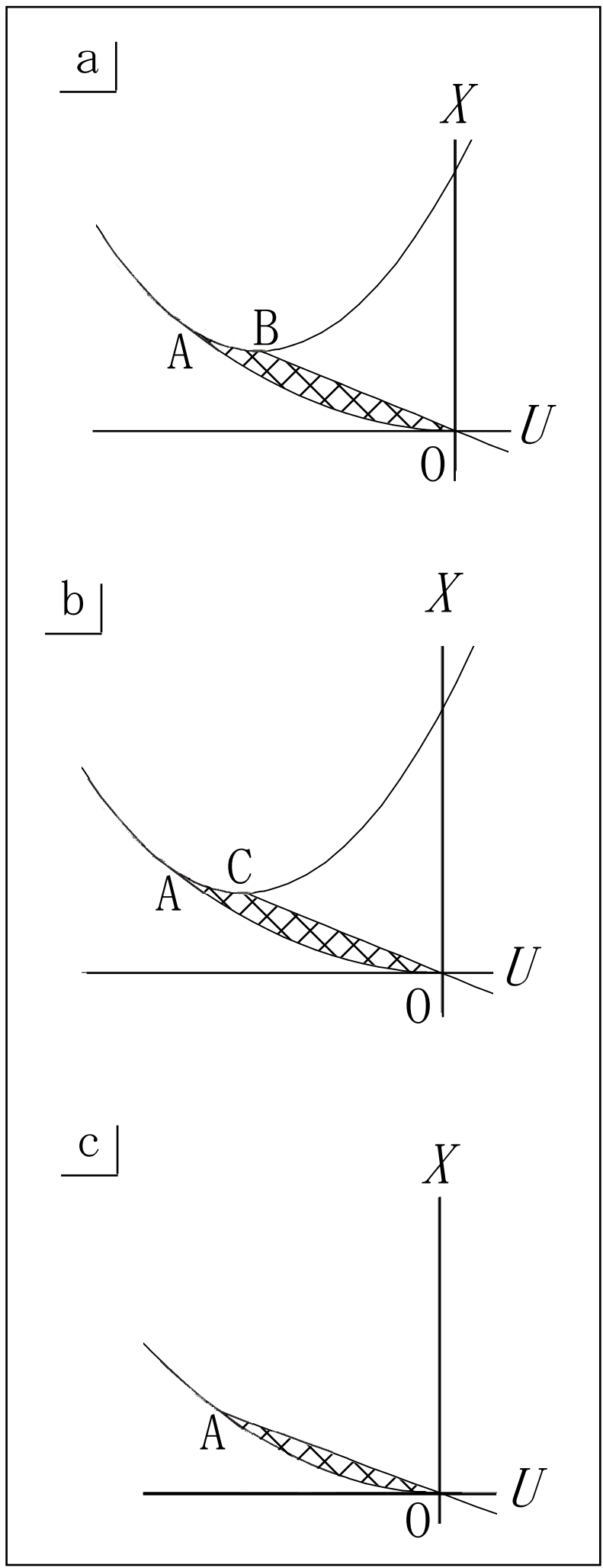
\item{Fig.~4.}
The same as Fig.~3 but for ($G$$<$$0$, $Q$$<$$0$).
     $\overline{OA}$ is $X$$=$$U^2$ and 
     the point $A$ is ($Q$, $Q^2$)
     through these figures.
a) : $Q$$<$$-1$, $0$$<$$F'$, \ 
     $\overline{AB}$ is $X$$=$$X_4(U,0)$.
     The point $B$ is ($\frac{1}{Q}$, $1$). \ 
b) : ($Q$$<$$-1$, $-3$$<$$F'$$<$$0$) or
     ($-1$$<$$Q$$<$$0$, $-3$$<$$F'$$<$$-3(1-Q^2)$), \ 
     $\overline{AC}$ is $X$$=$$X_4(U,F')$.
     The point $C$ is $\left( \frac{1+F'/3}{Q}, 1+F'/3 \right).$ \ 
c) : ($-1$$<$$Q$$<$$0$, $-3(1-Q^2)$$<$$F'$).
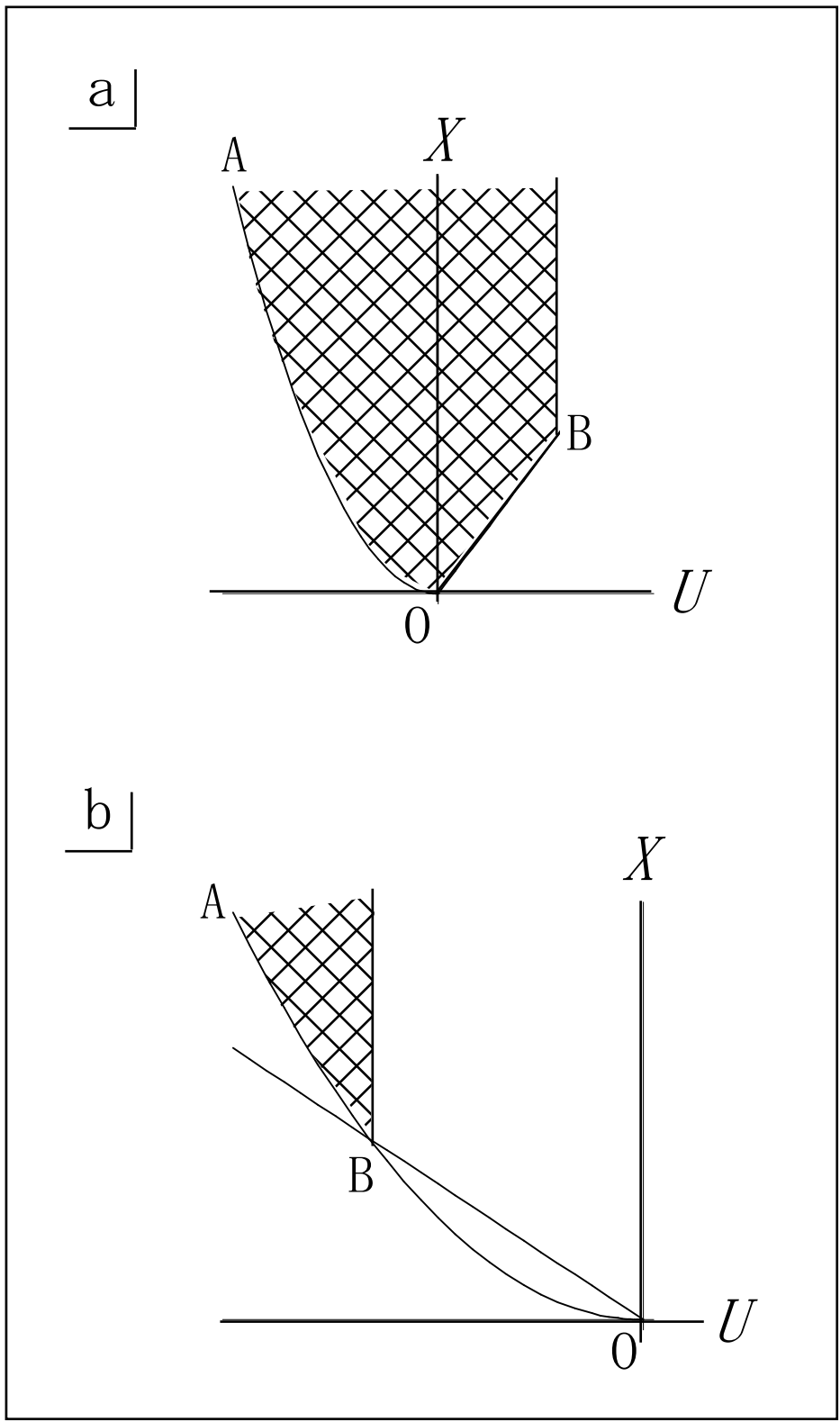
\item{Fig.~5.}
The region where both of {\bf M}- and {\bf A}-phase are defined 
and also $V_m$$<$$V_3$ is satisfied for $0$$<$$G$. \ 
     $\overline{OA}$ is $X$$=$$U^2$ and 
     the point $B$ is ($Q$, $Q^2$) in common.
a) : ($0$$<$$Q$$<$$1$, $-3(1-Q^2)$$<$$F'$). \  
b) : ($Q$$<$$0$, $-3$$<$$F'$).
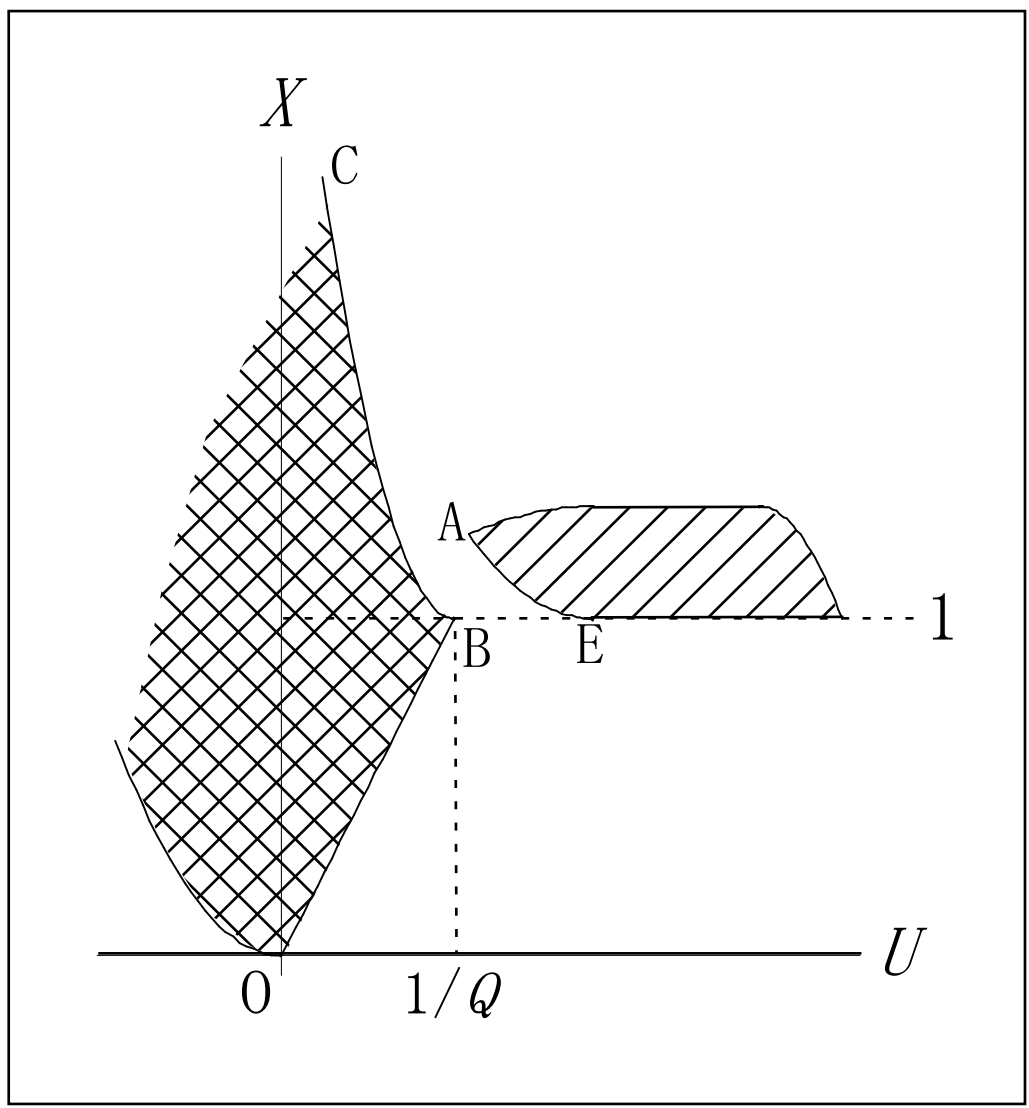
\item{Fig.~6.}
A illustration of $\Sigma_{m6}$ and $\Lambda$ 
for ($0$$<$$c$$<$$2$, $0$$<$$H'$), ($0$$<$$G$, $0$$<$$Q$) and $0$$<$$F'$.
This case corresponds to Fig.~1-a and Fig.~3-a. 
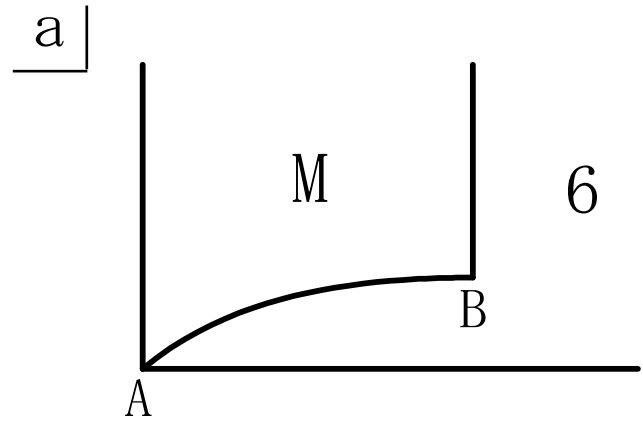
\item{Fig.~7.}
The phase diagrams 
{\bf a)} for No.~1-1, 
{\bf b)} for No.~1-4 and 
{\bf c)} for No.~1-5. \ 
$A$ : $\left( -1-H', -3 \right)$.   
$B$ : $\left( \frac{1}{Q}-H', 0 \right)$.  
$C$ : $\left( Q-H', -3 \right)$. 
$D$ : $\left( Q-H', 3(Q^2-1) \right)$.  
$\overline{AB}$ and $\overline{AD}$ : $F'_1(H)$.
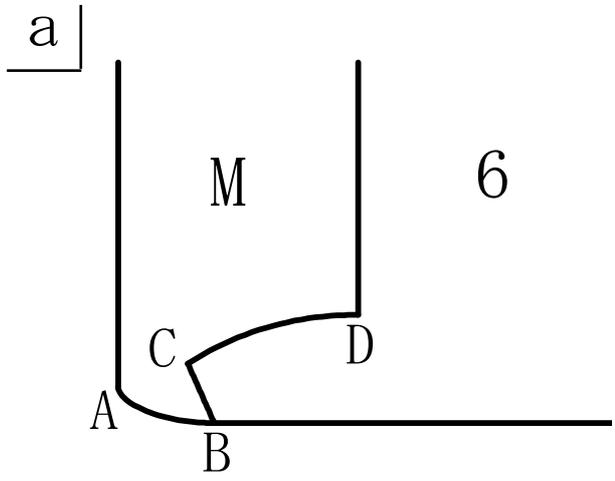
\item{Fig.~8.}
The phase diagrams 
{\bf a)} for No.~2-1, 
{\bf b)} for No.~2-2, 
{\bf c)} for No.~2-4, 
{\bf d)} for No.~2-5 and 
{\bf e)} for No.~2-6.\ 
$A$ : $\left( -1-H', (2-c)H' \right)$.  
$B$ : $\left( -\frac{1+c}{3}H', -3 \right)$.   
$C$ : $\left( H_a(c), F'_a(c) \right)$. 
$D$ : $\left( \frac{1}{Q}-H', 0 \right)$.  
$E$ : $\left( Q-H', * \right)$. 
$F$ : $\left( Q-H', 3(Q^2-1) \right)$. 
$\overline{AB}$ : $F'_I(H, c)$. 
$\overline{BC}$ : $F'_2(H, c)$. 
$\overline{CD}$ and $\overline{CG}$ : $F'_1(H)$.
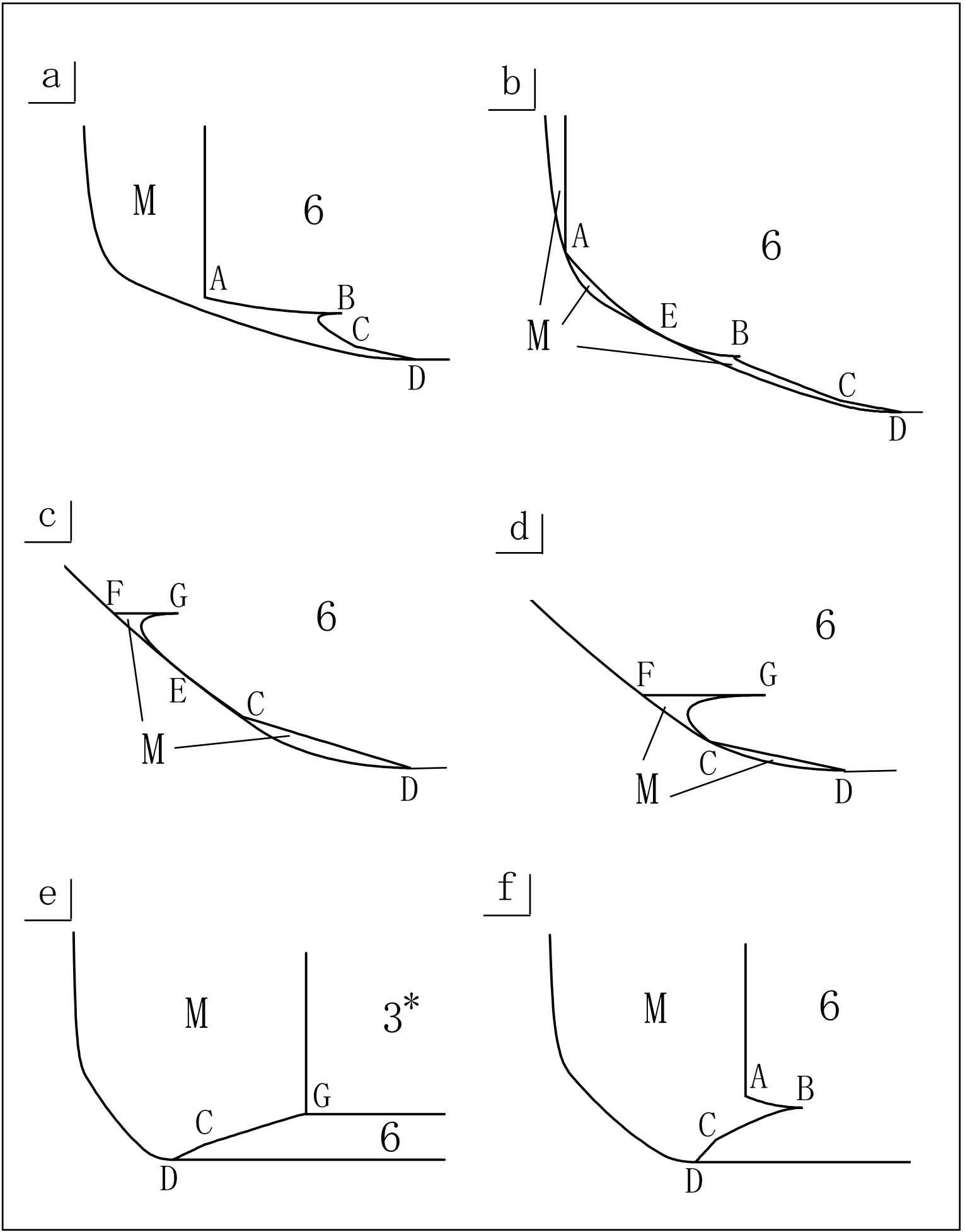
\item{Fig.~9.}
The phase diagrams 
{\bf a)} : $Q$$<$$-\frac{cH'+\sqrt{16+(cH')^2}}{4}$ and  
{\bf b)} : $-\frac{cH'+\sqrt{16+(cH')^2}}{4}$$<$$Q$$<$$-1$ for No.~3-1, 
{\bf c)} : $-1$$<$$Q$$<$$\frac{(c+2)H'-\sqrt{16+(c+2)^2H'^2}}{4}$ and 
{\bf d)} : $\frac{(c+2)H'-\sqrt{16+(c+2)^2H'^2}}{4}$$<$$Q$$<$$0$ for No.~3-2, 
{\bf e)} for No.~3-5 and
{\bf f)} for No.~3-6.\ 
$A$ : $\left( \frac{1}{Q}, \frac{(QcH')^2}{Q^2-1} \right)$
but in {\bf b)} $A$ is the intersection point of the two functions 
$F'_3(H, c)$ and $F'_{III}(H)$.
$B$ : $\left( \frac{cH'}{2}+\frac{1}{Q}, 0 \right)$. 
$C$ : $\left( H_a(-c), F'_a(-c) \right)$ 
but in {\bf d)} $C$ is the intersection point of the three functions 
$H_1(F', c)$, $F'_2(H, -c)$ and $F'_I(H, -c)$.  
$D$ : $\left( \frac{c-1}{3}H', -3 \right)$. 
$E$ : $\left( \frac{cH'}{2}+Q, 4(Q^2-1) \right)$ 
that is the tangent point of the functions $H_1(F', c)$ and $F'_{II}(H, c)$.
$F$ : $\left( \frac{cH'-\sqrt{1+3Q^2}}{2}, 3(Q^2-1) \right)$.
$G$ : $\left( \frac{cH'}{2}+Q, 3(Q^2-1) \right)$.
$\overline{AB}$ : $F'_3(H, c)$. 
$\overline{BC}$ and $\overline{CG}$ : $H_1(F', c)$. 
$\overline{CD}$ : $F'_2(H, -c)$.
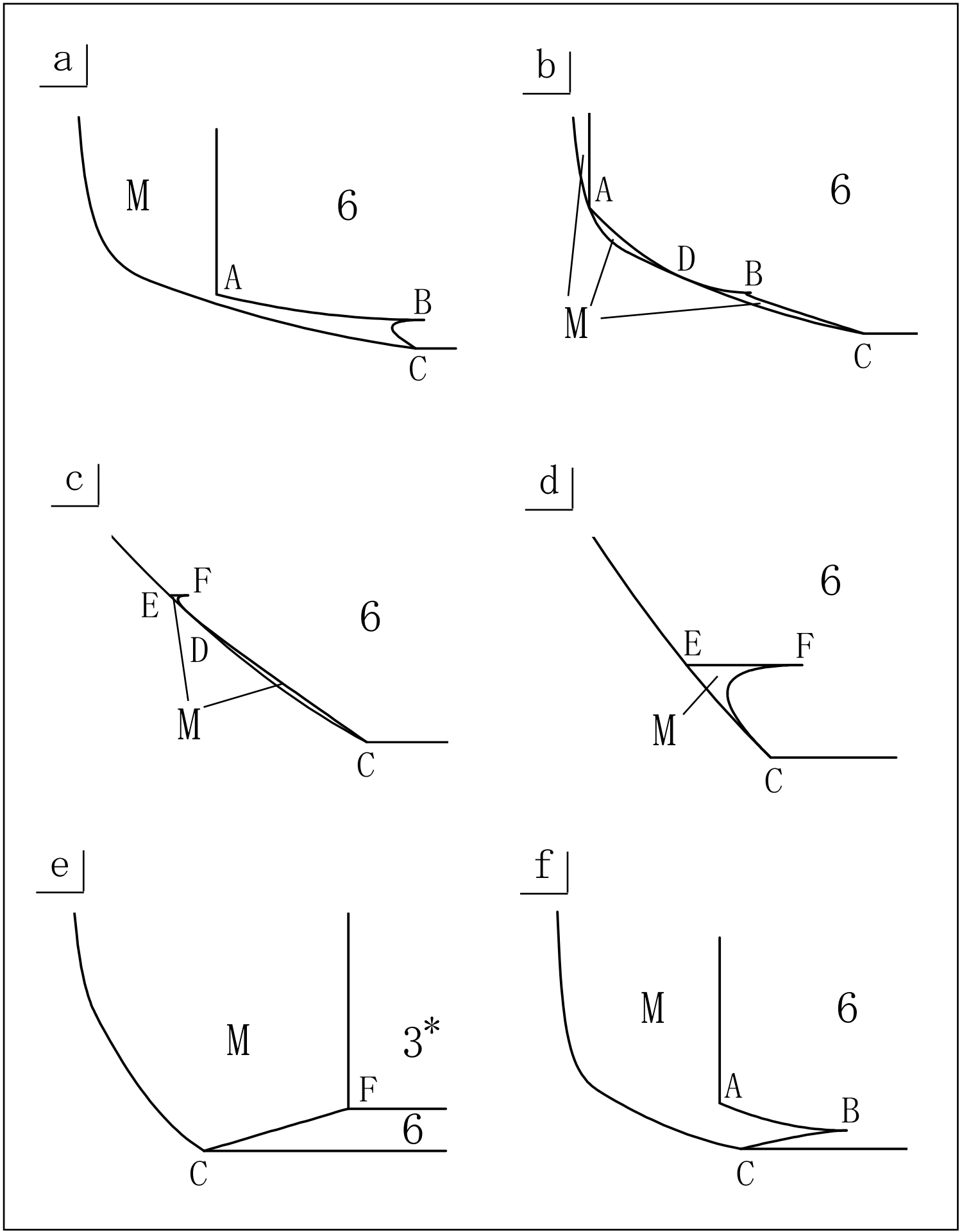
\item{Fig.~10.}
The phase diagrams 
{\bf a)} : $Q$$<$$-\frac{cH'+\sqrt{16+(cH')^2}}{4}$ and 
{\bf b)} : $-\frac{cH'+\sqrt{16+(cH')^2}}{4}$$<$$Q$$<$$-1$ for No.~4-1, 
{\bf c)} : $-1$$<$$Q$$<$$-\frac{1}{2}$ and 
{\bf d)} : $-\frac{1}{2}$$<$$Q$$<$$0$ for No.~4-2, 
{\bf e)} for No.~4-5 and
{\bf f)} for No.~4-6.\ 
$A$ : $\left( \frac{1}{Q}, \frac{(QcH')^2}{Q^2-1} \right)$. 
$B$ : $\left( \frac{cH'}{2}+\frac{1}{Q}, 0 \right)$.  
$C$ : $\left( \frac{cH'-1}{2}, -3 \right)$. 
$D$ : $\left( \frac{cH'}{2}+Q, 4(Q^2-1) \right)$ 
that is the tangent point of the functions $H_1(F', c)$ and $F'_{II}(H, c)$.  
$E$ : $\left( \frac{cH'-\sqrt{1+3Q^2}}{2}, 3(Q^2-1) \right)$.  
$F$ : $\left( \frac{cH'}{2}+Q, 3(Q^2-1) \right)$.  
$\overline{AB}$ : $F'_3(H, c)$. 
$\overline{BC}$ and $\overline{CF}$ : $H_1(F', c)$ 
that is replaced by $F'_{+}(H, c)$ in {\bf e)} and {\bf f)}.
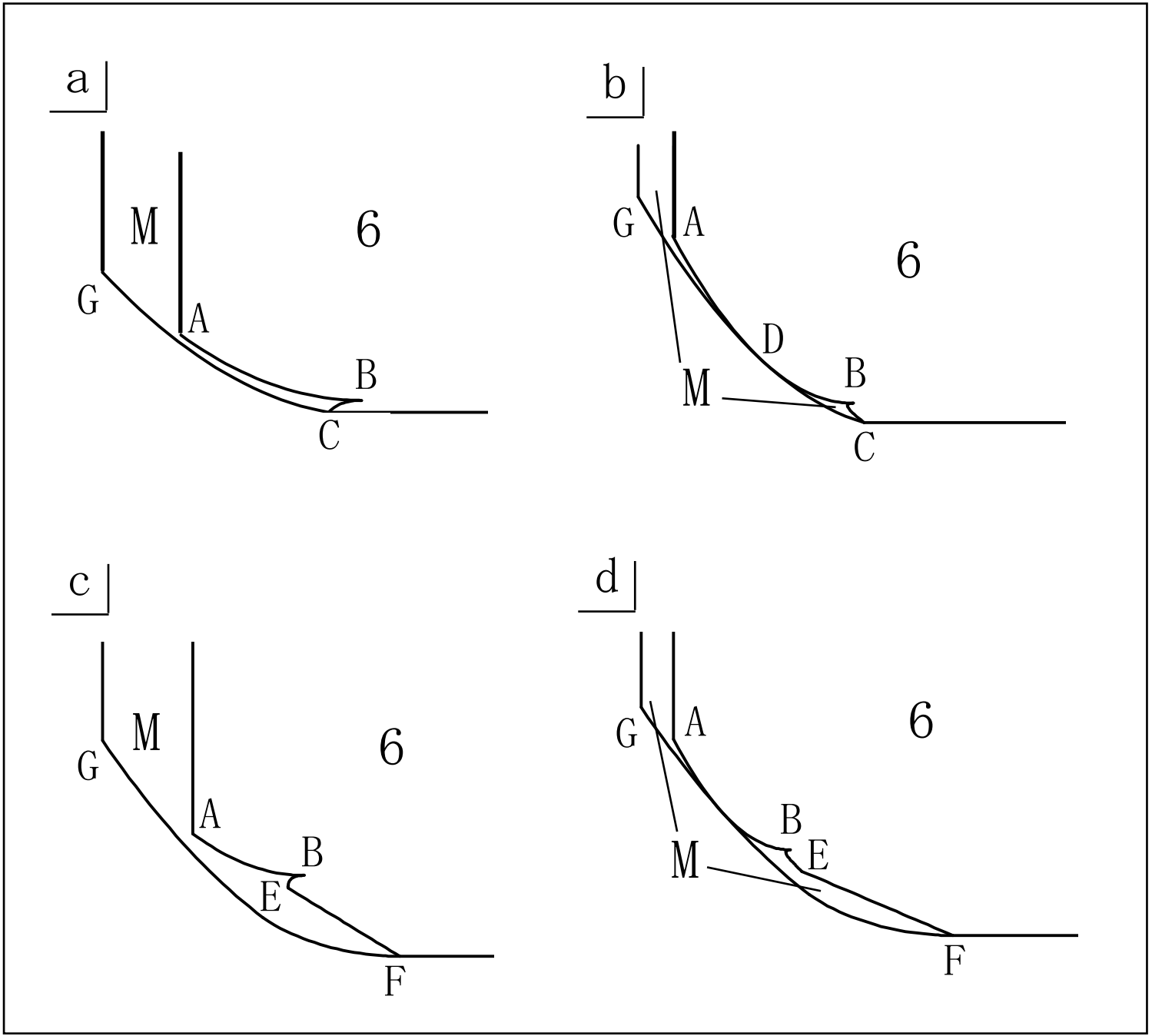
\item{Fig.~11.}
The phase diagrams 
{\bf a)} and 
{\bf b)} for No.~5-1, 
{\bf c)} and 
{\bf d)} for No.~6-1.\  
$A$ : $\left( \frac{1}{Q}-H, \frac{\left\{ Q(c-2)H' \right\}^2}{Q^2-1} \right)$.  
$B$ : $\left( \frac{1}{Q}-\frac{cH'}{2}, 0 \right)$. 
$C$ : $\left( -\frac{cH'+1}{2}, -3 \right)$. 
$D$ : $\left( Q-\frac{cH'}{2}, 4(Q^2-1) \right)$ 
that is the tangent point of the functions $F'_3(H, -c)$ and $F'_{II}(H, -c)$. 
$E$ : $\left( H_a(c), F'_a(c) \right)$. 
$F$ : $\left( -\frac{c+1}{3}H', -3 \right)$. 
$G$ : $\left( -1-H', \left\{(c-2)H'-2 \right\}^2-4 \right)$.  
$\overline{AB}$ : $F'_3(H, -c)$. 
$\overline{BC}$ and $\overline{BE}$ : $H_1(F', -c)$.  
$\overline{EF}$ : $F'_2(H, c)$.
\end{description}

\begin{thebibliography}{99}
%
\bibitem{NYS} E.Nakano, T.Suzuki and H.Yabu, in preparation to publication.
%
\bibitem{CSC} see references in \cite{NYS}.
%
\bibitem{VOL} G.E.Volovik, ``EXOTIC PROPERTIES OF SUPERFLUID ${}^3$He''
(World Scientific Pub. Co. Ote. Ltd., Singapore, 1992) 
%
\bibitem{GS} J.Goldstone, Nouvo Cim. 19(1961) 154.
%
\end{thebibliography}
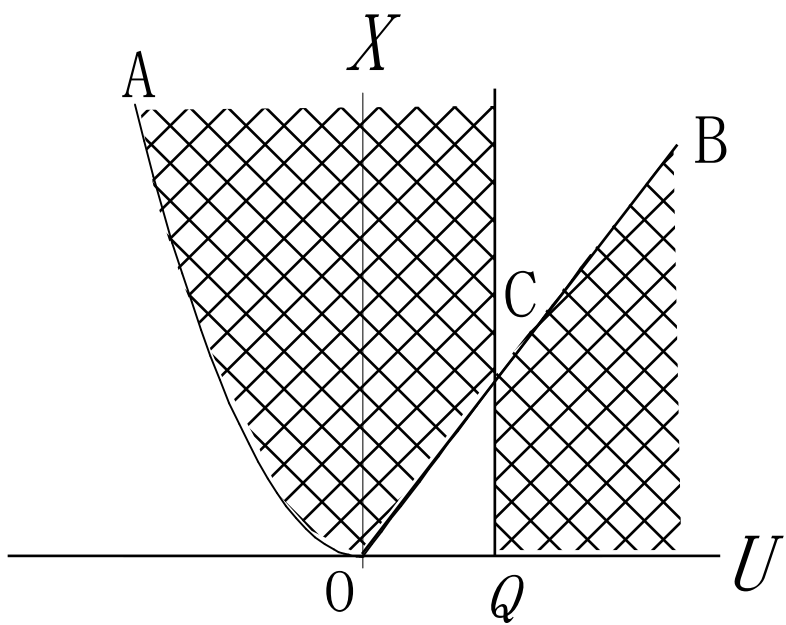
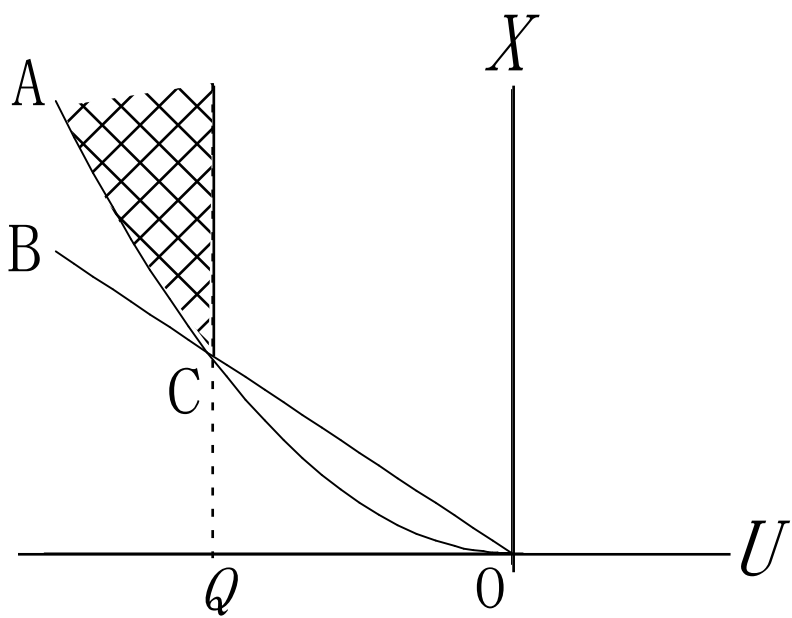
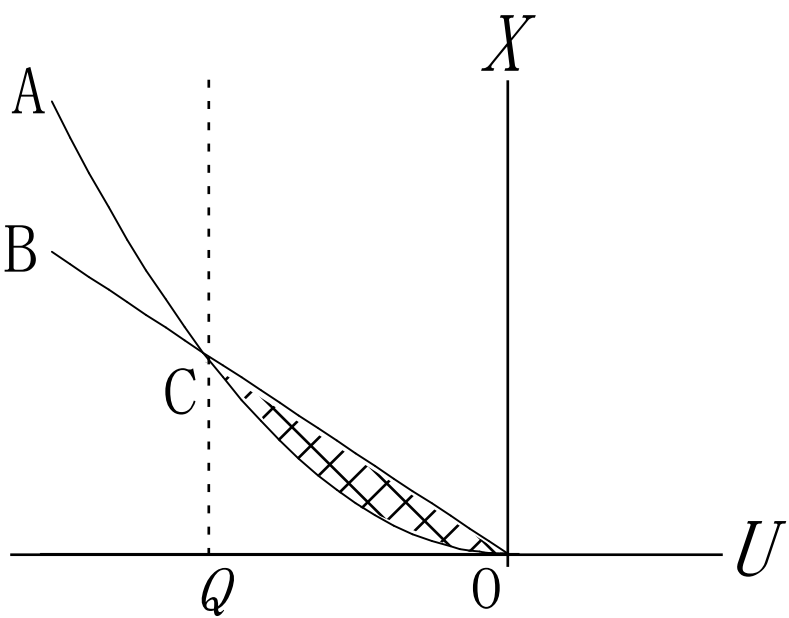
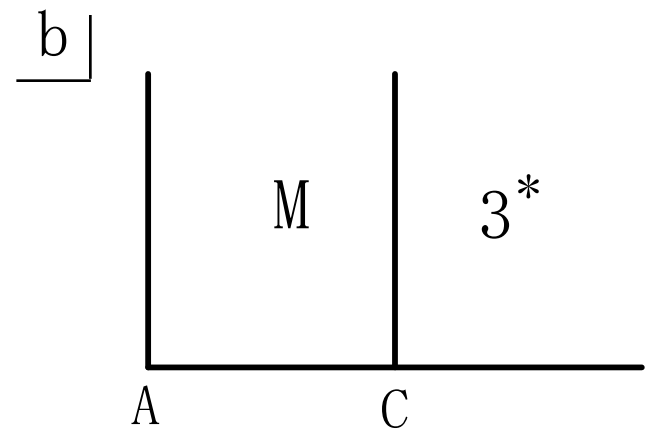
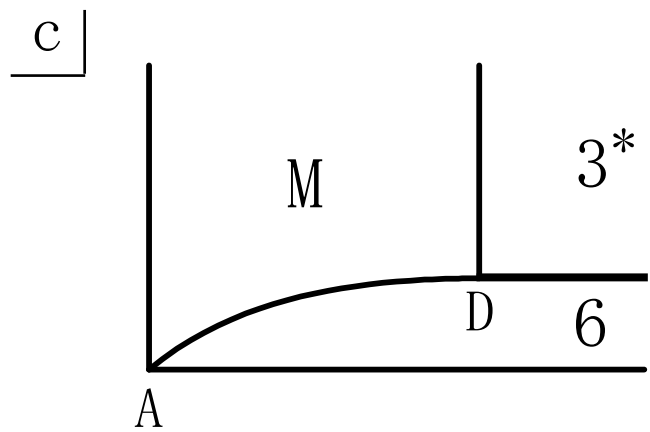
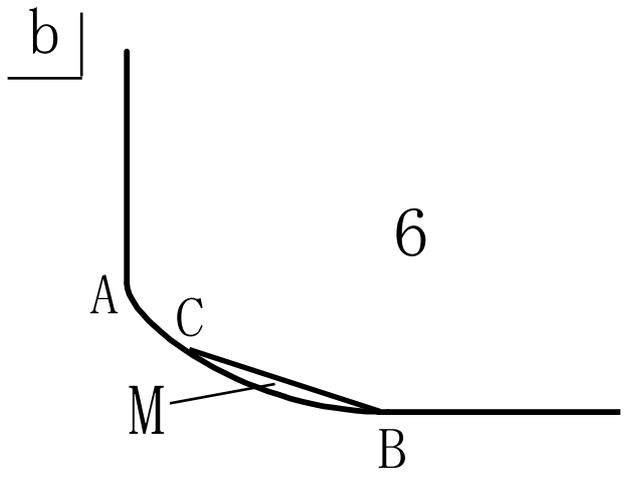
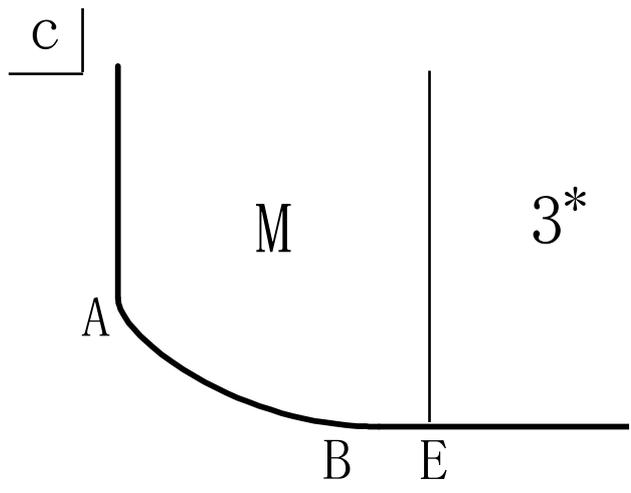
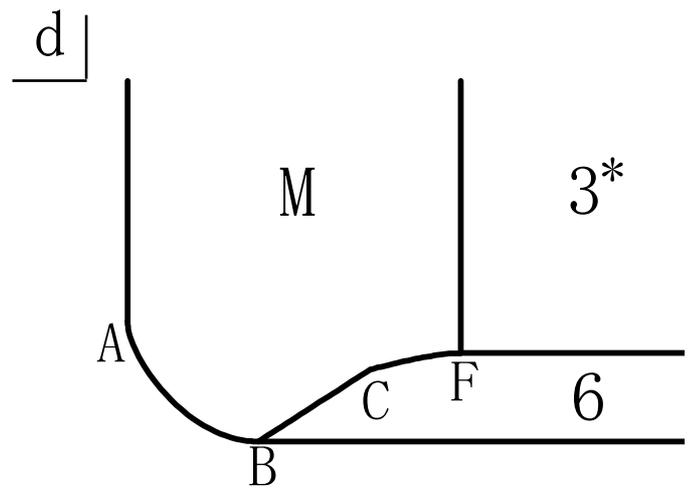
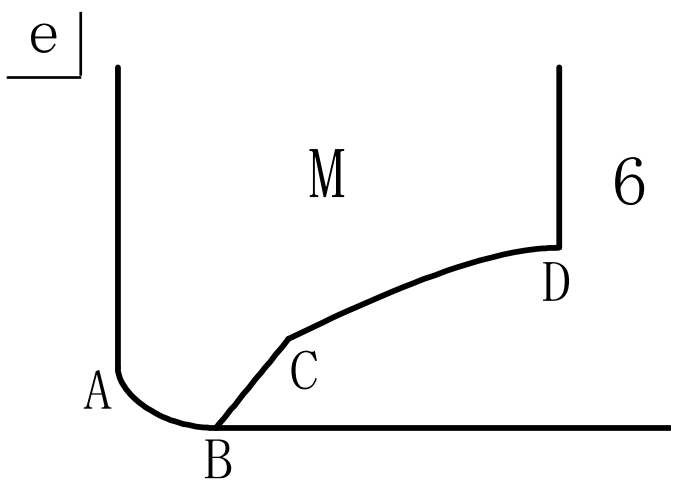
\end{document}